\documentclass[11pt]{article}
\usepackage[utf8]{inputenc}
\usepackage{amsmath}
\usepackage{graphicx}
\usepackage[top=3cm, bottom=3cm, left=2cm, right=2cm]{geometry}

\title{Local Algorithms for Graphs}
\author{
David Gamarnik, MIT Sloan School of Management, United States\\
Mathieu Hemery, Univ.\ Grenoble Alpes, LIPhy, France\\
Samuel Hetterich, Goethe University, Mathematics Institute, Germany. \\
}
\date{20 november 2013}

\begin{document}

\maketitle{}

\textit{These are the notes from the lecture by David Gamarnik given at the autumn school “Statistical Physics, Optimization, Inference, and Message-Passing Algorithms”, which took place at Les Houches, France, from September 30th to October 11th 2013. The school was organized by Florent Krzakala from UPMC \& ENS Paris, Federico Ricci-Tersenghi from La Sapienza Roma, Lenka Zdeborova from CEA Saclay \& CNRS, and Riccardo Zecchina from Politecnico Torino.}

\tableofcontents

\section{Introduction}
We are going to analyze local algorithms over sparse random graphs. These algorithms are based on local information where local regards to a decision made by the exploration of a small neighbourhood of a certain vertex plus a believe of the structure of the whole graph and maybe added some randomness. This kind of algorithms can be a natural response to the given problem \cite{Linial} or an efficient approximation such as the Belief Propagation Algorithm \cite{MM}.

\subsection{The independent set problem}
The independant set problem,  a long running problem in the history of graph theory \cite{Cook, Coja_11}, can be define as follow.
Given a Graph $G$ with vertex set $V$ and edge set $E$, we say that a set $I\subset V$ is an independent set (and abbreviate it i.s.) of $G$ if for all $v,v' \in G$ we have $(v,v')\notin E$. Let $\mathcal{I}_G = \left\{I\subset V|I \text{ is an independent set of }G \right\}$ be the set of all independent sets of $G$. We can think of algorithm which counts the number of all independent sets of $G$ or we could ask for
$$\max_{I\in \mathcal{I}_G} |I|$$
a largest independent set of $G$. If $G$ is a weighted graph such that we have a weight $w(v)$ on every vertex $v\in V$ we can ask for 
$$
\max_{I\in \mathcal{I}_G} w(I)
$$ 
an independent set of $G$ with the largest weight $w(I) = \sum_{v\in I} w(v)$. Further on this lecture we are going to see an algorithm which counts the number of independent sets $|\mathcal{I}_G|$ for a special class of graphs.

It is still surprising that for a good class of problems local algorithms provide a good prediction, a near optimal or even best we could hope of solution.\\
Let us now focus on the problem of counting the number of all independent sets of a graph and denote this number by $Z$. In the case of sparse graphs this is actually exponential in the number of vertices. Let us also introduce a parameter $\lambda > 0$ and a partition function 
$$Z_\lambda=\sum_{I\in \mathcal{I}_G} \lambda^{|I|}$$
such that $Z_\lambda =Z$ when $\lambda=1$. In this case we just write $Z_G$ beeing the number of all independent sets of $G$. In statistical physics it is known as the partition function of the \emph{Hard-Core-Model} with parameter $\lambda$ \cite{Berg-Steif}.  \\
Let us now fix $\lambda =1$ and $G$ be a graph with bounded degree such that the degree of every vertex of $G$ has degree less or equal to $d$. If we choose an independent set $I$ of $G$ uniformly at random out of $\mathcal{I}_G$ (which is algorithmically highly non trivial) we compute $\Pr[v\notin I]$ the probability that a certain vertex $v\in V$ is not in $I$. It is easily verified that 
\begin{align}
\Pr_G[v\notin I] = \frac{Z_{G-v}}{Z_G} = \frac{\#\text{indep. sets without }v}{\# \text{independent sets}}
\end{align}
This is equivalent to 
\begin{align}
Z_G = \Pr_G[v\notin I]^{-1} \cdot Z_{G-v}
\end{align}
By ordering the vertices in $V$ with $|V|=n$ we have the recursion
\begin{align}
Z_G = \Pr_G[v_1\notin I] \cdot Z_{G-v_1} = \Pr_{G-v_1}[v_1\notin I]^{-1} \cdot \Pr[v_1\notin I]^{-1} \cdot Z_{G-v_1-v-2} \\ =\ldots = \prod_i  \Pr_{G_{i-1}}[v_i\notin I]^{-1}
\end{align}
where $G_i=G-v_1-v_2-\ldots-v_i$.\\ 
Notice that this quantity is invariant under changing the order of the vertices which are deleted. By estimating this probabilities we gain an estimate of the partition function. We will see that this can be done exactly for a certain class of graphs by just exploring a small neighbourhood in the thermodynamic limit as $n$ tends to infinity and gives a good approximation for finite $n$.

\section{Power of the local algorithms}

\subsection{Hosoya index}
A second example to which we can adopt local algorithms to estimate the partition function which is defined analogously to independent sets is the problem of counting the number of matching on a graph, also known as Hosoya index \cite{Karp}. A matching is a set $M\subset E$ such that for any two edges $(v_1,v_2), (v_3,v_4) \in M$ all $v_i$ are pairwise different. The considered counting problem is in $\# P$ for a sparse random graph.  

Let us define the girth length $g(G)$ of a graph as the length of a shortest cycle of $G$. We say a sequence of graphs $G_n$ with bounded maximal degree $d$ is locally tree-like if $g(G_n)$ tends to infinity as $n$ grows. Notice that sparse random regular graphs are only approximately locally tree-like by occurrence of a constant number of short cycles. 

\subsection{Theorem}
\label{theo_d_reg_l_t_l}
Let $G_n$ be a sequence of a $d$-regular locally tree-like graphs of n nodes. If 
\begin{align}
\lambda \leq \lambda_c(d)= \frac{(d-1)^{d-1}}{(d-2)^d}
\end{align}
we have
\begin{align}
\lim_{n\rightarrow \infty} \frac{\log Z_{G_n}}{n} = -\log (p^{\frac{d}{2}}(2-p)^{\frac{d-2}{2}})
\end{align}
for $d=2,3,4,5$ where $p$ is the unique solution of $p=(1+\lambda p^{d-1})^{-1}$.

\subsection{Remark}
\begin{itemize}
\item The statement is probably wrong for $d>5$. 
\item The statement can be generalized to sparse random regular graphs. 
\item We can not hope to get such a general bound for the maximal independent set since we have an upper bound on it for random $d$-regular graphs which is $0.459n$ and trivially the maximal i.s. of a random $d$-regular bipartite graph is at least $0.5n$. We conclude that the considered class of graphs in  Theorem \ref{theo_d_reg_l_t_l} is to broad to get a similar result on maximal independent sets.
\item Theorem \ref{theo_d_reg_l_t_l} can be stated for matchings and without the restrictions on $\lambda$ and $d$.
\end{itemize}

\subsection{Proof}
\begin{figure}[ht]
\centering
\includegraphics[width=0.5\textwidth]{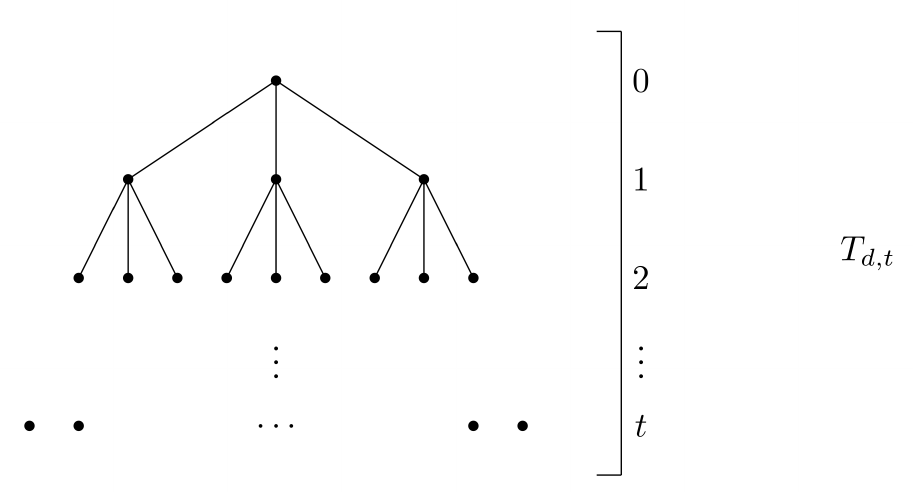}
\caption{Sketch of a $d$-regular tree with depth $t$. \label{fig:graph}}
\end{figure}

Consider a rooted $d$-regular tree where the root denoted by $v$ has $d$ children and every child has again $d$ children as depicted in Fig. \ref{fig:graph}. 

If this tree has depth $t$ it is uniquely defined and we refer to it as $T_{d,t}$. We introduce the shorthand $Z_{T_{d,t}}=Z_t$ if $d$ is specified. Let us denote $Z_G[B]$ the number of independent sets which satisfy condition $B$.\\
We find that 
\begin{align}
Z_t= Z_t[v\notin I] + Z_t[v\in I]= Z_t[v\notin I]+ (Z_{t-1}[v'\notin I])^d
\end{align}
where $v'$ is a child of $v$ which is equivalent to 
\begin{align}
\frac{Z_t}{Z_t[v\notin I]}= 1 + \frac{(Z_{t-1}[v'\notin I])^d}{Z_t[v\notin I]} = 1 + \left(\frac{(Z_{t-1}[v'\notin I])}{Z_{t-1}}\right)^d
\end{align}
We have already interpreted the quantity on the left hand side as the probability that $v$ is not contained in a uniformly at random chosen independent set of $T_{d,t}$ which we denote by $P_t$.\\
Thus we get 
\begin{align}
P_t=(1+P_{t-1}^d)^{-1}
\end{align} 
which is a recursion that hopefully converges to a fixed point.\\
We introduce the function $f(x)=(1-x^d)^{-1}$ defined on $[0,1]$ which is decreasing. Let us analyze the second iteration of $f$ namely $f^{(2)}=f\circ f$ which is clearly increasing on $[0,1]$ and well defined since $f([0,1])\subset [0,1]$.\\
Thus we get 
\begin{align}
0 &\leq f^{(2)}(0) \leq f^{(4)}(0) \leq f^{(6)}(0) \leq \ldots \leq f^{(2n)}(0)\leq \ldots \rightarrow x_*\\
1 &\geq f^{(2)}(1) \geq f^{(4)}(1) \geq f^{(6)}(1) \geq \ldots \geq f^{(2n)}(1)\geq \ldots \rightarrow x^*
\end{align}
If we are lucky we get $x_*=x^*$ having a unique fixed point in $[0,1]$. To check this, we plot the monotonic increasing function $f^{(2)}$ and check how many times it intersects the identity. If this is only once the case we have a unique fixed point. It turns out that this is the case for $d=1,2,3,4$ and analogously if $\lambda \leq \lambda_c(d)= \frac{(d-1)^{d-1}}{(d-2)^d}$.\\
If it exists let us define the unique fixed point as $\lim_{t\rightarrow \infty} P_t = P^*$.\\
Consider the tree $T_{d,t}$ and define any boundary condition $B$ by fixing every vertex of depth $t$ on being in an independent set or out. Let $B_{in}$ (resp. $B_{out}$) be the event that all vertices of depth $t$ are in (resp. not in) the independent set.
It turns out that we have for any boundary condition $B$
\begin{align}
 \Pr[v\notin I|B] \leq \max\left\{\Pr[v\notin I|B_{in}],\Pr[v\notin I|B_{out}]\right\}
\end{align}
Moreover we have either 
\begin{align}
 \Pr[v\notin I|B_{out}] \leq P_{t-1}
\end{align}
or 
\begin{align}
 \Pr[v\notin I|B_{in}] \leq P_{t-2}
\end{align}
and thus $P^* \in [\min\left\{P_{t-1},P_{t-2}\right\},\max\left\{P_{t-1},P_{t-2}\right\}]$.\\
On the whole graph we obtain
\begin{align}
 \Pr_{G}[v\notin I] = \sum_B\Pr_t[v \notin I|B]\cdot \Pr[B|G] \sim P^*\sum_B \Pr(B|G)=P^*
\end{align}
which is true as long as we don't have any loop. We now have two last corrections to complete the proof. The first one is a small one,
in a regular tree the root has a degree $d-1$ while all the other nodes has degree $d$. But this is easily corrected by introducing:
\begin{align}
P^*_{d-1} = \frac{1}{1+{P^*_{d-1}}^{d-1}},
\end{align}
for the first recursion step.

The other problem is more serious. For yet, we want to use the recursive formula $Z_G=p^{-1}Z_{G-v}$ by removing the node one by one. But if we do so, $G-v$ is no longer a regular graph.
The trick is to remove two vertices $u$ and $v$, and to connect their children in order to preserve the regularity property. But be carefull, $u$ and $v$ must be chosen sufficiently far away in order to preserve also the locally tree structure.

Let us note $G_{n-2}=G_n-u-v+e_1+\ldots+e_d$ where $\{e_i\}$ denotes the new link between the children of $u$ and $v$ and $H=G_n-u-v$.
We want to find the relation between $Z_{G_n}$ and $Z_{G_{n-2}}$ as a function of $p$. We have $Z_{G_n}=(2-p)^2 Z_H$. So let us look at the effect of adding a new link, $e_1$ for example.
\begin{align}
\frac{Z_{H+e_1}}{Z_H}=\frac{\#\text{indep. sets without }i\text{ and }j}{\#\text{indep. sets in }H}\simeq1-(1-p)^2,
\end{align}
because of the independance of $u$ and $v$ which ensure that $\Pr(u,v \in I)=\Pr(u \in I) \Pr(v \in I)=(1-p)^2$.

As we repeat the operation $d$ times we have
\begin{align}
\frac{Z_{G_{n-2}}}{Z_H}=\frac{H+e_1+\ldots+e_d}{H}=(1-(1-p)^2)^d=(2p-p^2)^d,
\end{align}
and at the end of the day
\begin{align}
\frac{Z_{G_{n}}}{G_{n-2}}=\frac{1}{p^d(2-p)^{d-2}}.
\end{align}

As this formula is true $n/2-o(n)$ times, we can write $Z_{G_n}\sim(p^d(2-p)^{d-2})^{-n/2}$ that is
\begin{align}
\frac{\log Z_{G_{n}}}{n}\rightarrow-\log{\left(p^{\frac{d}{2}}(2-p)^{\frac{d-2}{2}}\right)}.
\end{align}

\section{Incursion in a loopy land}

We are now interested in the case where $G$ is no more locally tree-like, nor regular. We just ask that the degree of $G$ is bounded
that is: $\forall i\in V,deg(i)\leq D$ where $deg(i)$ denotes the
degree of the node $i$ and $D$ is an integer.

We still try to compute the partition function $Z_{\lambda}$ but
will set by default $\lambda=1$ and we'll use the notation $Z(G)=Z_{G}(1)$, to emphasize the graph on which we're working on.
Of course the formula
\begin{equation}
P_{G}(u\notin I)=\frac{Z(G-u)}{Z(G)},\label{eq:recursion}
\end{equation}
still remain. Let us note $v_{0}$ the current node and $v_{1},\ldots,v_{d}$
($d=deg(v_{0})$) his neighbours and define $G_{i}=G-v_{0}-\ldots-v_{i}$.

\paragraph{Exercise:}

Check that formula 
\begin{equation}
P_{G}(v_{0}\notin I)=\frac{1}{1+{\displaystyle \prod_{i=0}^{d}}P_{G_{i-1}}(v_{0}\notin I)},
\end{equation}
is correct.

This define the computational tree for which $G$ is the root and
$G_{i}$ the children. On this tree we compute on each node the probability
of $v_{i}$ to be out of the independant set. And we can proove that
their exist some value of the parameters $\lambda,D$ for which the
correlation are decreasing on the computational tree. Indeed this
is a corrolar from the contraction property which state that when
defining a function $f(x_{1},\ldots,x_{n})=\left(1+\lambda x_{1}\ldots x_{n}\right)^{-1}$,
if $\tilde{x_{i}}$ is an approximation of $x_{i}$ up to $\delta$.
It follows that $f(\tilde{x_{1}},\ldots,\tilde{x_{n}})$ is also an
approximation of $f(x_{1},\ldots,x_{n})$ up to $\rho\delta$ where
$\rho\leq1$.

\section{The monomer-dimer entropy}

$B_{n}^{d}$ is the regular hypercube of size $n$ and dimension $d$.
It can be seen as a graph with $n^{d}$ vertex. The monomer-dimer
entropy result from the number of partial matching you can find in
such a graph. As usual, $Z(B_{n}^{d})$ is the value of the partition
function for $\lambda=1$. And if we don't care about assigning energy
to the different configuration, the entropy is define as $H=\log\left(Z(B_{n}^{d})\right)$.
Which grow according to $n^{d}$. It is so tempting to define the
intensive entropy, and by taking the limit $n\rightarrow\infty$,
we pose:
\begin{align}
h(d)=\lim_{n\rightarrow\infty}\frac{\log\left(Z(B_{n}^{d})\right)}{n^{d}},
\end{align}
the entropy of the MD-model (see for exemple \cite{Hammersley} for an hisotrical
paper and \cite{Friedland} for more recent work).

How can we compute such a quantity? A physicist will propose to use
transfer matrix and this has been done for $d=3$ giving the bounds:
\begin{align}
0.785\leq h(3)\leq0.786.
\end{align}
But it can be shown that computing $h$ with an error $\epsilon$
scale as $\exp(o\left(\frac{1}{\epsilon}\right)^{d-1})$ which fastly
became intractable.

By using the cavity method on the graph, adding one vertex at a time
the bounds have been increase to
\begin{align}
0.78595\leq h(3)\leq0.785976.
\end{align}

\paragraph{Remark:}
It's far more difficult to compute the dimer model who correspond
to the perfect matching, than the monomer-dimer which by allowing
gaps between the edge correspond only to a partial matching.

\section{Limits of local algorithms}

To give an insight into the limits of what a local algorithm can do,
we'll focus on the case of random graph and especially on Erdos-Renyi
graph ($G\left(n,\frac{d}{n}\right)$) and regular graph ($G_{d}(n)$).
We still remain into the independant set problem but are now trying
to find the maximal indepandant set $I_{n}^{\star}$.

Of course, this quantity scale lineraly with $n$ and so for Erdos-Renyi
graph, we introduce
\begin{align}
\alpha(d)=\lim_{n\rightarrow\infty}\frac{I_{n}^{\star}}{n}.
\end{align}
It can be shown that
\begin{align}
\alpha(d)\sim\frac{2\log d}{d}.
\end{align}
But it seems that the local algorithms get always stucks around the
threshold $t_{d}=\frac{\log d}{d}$ which let us a factor $2$ of
improvement \cite{Gamarnik_14}. Let us describe a very naive algorithm, the so called
``greedy algorithm'': choose a random node $v$ in $G$ and put
it in the indepandent set $I$, then erase all of his neighbour and
cycle as long as their exist a node to put in $I$. This (stupid)
algorithm is always stuck around $t_{d}$. But no local algorithm can
do beter!

Let's first define more precisely the notion of ``local'' algorithm.

For $G$ a regular random graph of degree $d$. A ``local''
algorithm can always be represented by a function $f_{r}:\left[0,1\right]^{\left|T_{d,r}\right|}\rightarrow\left\{ 0,1\right\} $
and a realisation of the algorithm is given by a set of weight $W=\left(w_{1},\ldots,w_{n}\right)$
set on the nodes of the graph. Where $T_{d,r}$ is the regular tree
of degree $d$ and depth $r$.

$f_{r}$ should be seen as a decision function based on the neighbourhood
of each nodes which is locally tree-like up to a depth $r$. It then
answers a value $0$ or $1$ assigning for example the node into the
independant set or not.

Can such a function give a correct realisation of the IS-problem?
Obviously yes, let $r=1$ and $f_{r}$ be $1$ if the weight of the
current node is higher than this of all his neighbour and $0$ otherwise.
This define a valid independant set, even if it'll surely not be the
maximal one.

The greedy algorithm can also be define in term of such a function.
For exemple, $W$ represent the order in which the node are chosen
by the greedy algorithm. When running $f_{r}$ on the graph, you have
to check that no neighbour have been check before the current node.
If it is the case everything is OK, but if another node could have
been check before, you must also check this new node and so on and
so forth. It's not obvious that this process will have an end before
exploring the whole graph. But remark that each node increase a bit
the value $w$ that the neighbour should beat to be chek. This increasing
value ensure that the number of path goes as $\frac{1}{r!}d^{r}$
where $r=\frac{2\log n}{\log\log n}$ and so stay in finite size.
This kind of property is often refered as influence resistance.

So, why is not possible that such a local algorithm does better than
$t_{d}$?

The analysis show that above this threshold, the independant sets
satisfies a clustering property. Roughly speaking, that means that
for $I$ and $J$ two differents independant set on $G$. Either $\left|I\cup J\right|$
is small, either it's large. This kind of feature has been discussed and proved in various geometry see \cite{Tal10, MMZ05, ACORT11}.

Let's now $f_{r}$ be the perfect local algorithm such that it produce
independant set and $\alpha(f_{r})\sim\alpha(d)$. We can run it twice
on the graph $G$ and obtain $f_{r}(U)=I$ and $f_{r}(V)=J$ two differents
independant set on $G$. As these are two independant realisation,
we have $\left|I\cup J\right|\sim\alpha(f_{r})^{2}=o\left(\frac{1}{d^{2}}\right)$
which is a very tiny fraction of the graph but we can generate a family
of such independant set by running succecively the function on the
ensemble $W_{t}=(u_{1},\ldots,u_{t},v_{t+1},\ldots,v_{n})$. And as
$f_{r}$ only depend on local variable to make his decision, that
also means that the succesive results $I_{t}=f_{r}(W_{t})$ only differ
locally, that means $\left|I_{t}\diamond I_{t+1}\right|=o(1)$ (where
$I\diamond J$ denotes the nodes which change between $I$ and $J$)
but that imply also that for a value of $t$ we must have $\left|I_{t}\cup I\right|$
which fall into the forbidden area, which is impossible according
to the clustering property.

\bibliographystyle{unsrt}

\end{document}